\documentclass[journal, a4paper, final]{IEEEtran}
\usepackage[english]{babel}
\usepackage{graphicx}
\usepackage{amsmath}
\usepackage{amsfonts}
\usepackage{amssymb}
\usepackage{hyperref} 
\hypersetup{
	colorlinks=false, 
	pdfborder={0 0 0} 
}

\usepackage{booktabs}
\usepackage{siunitx}
\usepackage{amsfonts,eucal,bm}
\usepackage{color}
\usepackage{pgfplots}
\usepackage{mleftright} 
\usepackage{amsmath}
\usepackage{amssymb}
\usepackage{multirow}

\ifCLASSOPTIONcompsoc
\usepackage[caption=false,font=normalsize,labelfont=sf,textfont=sf]{subfig}
\else
\usepackage[caption=false,font=footnotesize]{subfig}
\fi
\pgfplotsset{compat=newest,
	HardPlotStyle/.style={
		xlabel={$p$ of $\mathrm{BSC}$},
		xmax = 0.099,
		xmin = 0.010000,
		ylabel={$\mathrm{WER}$},
		ymax = 1,
		xmajorgrids,
		ymajorgrids,
		grid style=dashed,
		legend pos = south east,
		line width=1pt,
		every axis y label/.style={at={(current axis.north west)},above left=3mm},
	},
	SoftPlotStyle/.style={
		xlabel={$\frac{E_b}{N_0}$ in dB},
		xmax = 3.000000,
		xmin = 1.000000,
		xtick={1, 1.5, 2, 2.5, 3, 3.5, 4, 4.25, 4.5},
		ylabel={$\mathrm{WER}$},
		ymax = 1,
		xmajorgrids,
		ymajorgrids,
		grid style=dashed,
		legend pos = south west,
		line width=1pt,
		every axis y label/.style={at={(current axis.north west)},above left=3mm},
	},
	SizeStyleLarge/.style={
		width = \linewidth,
		label style={font={\footnotesize \color{white!15!black}}},
		tick label style = {font = \footnotesize},
		legend style={font = \footnotesize,legend cell align=left, align=left, draw=white!15!black},
	},
	SizeStyleSmall/.style={
		width = \linewidth,
		label style={font={\footnotesize \color{white!15!black}}},
		tick label style = {font = \footnotesize},
		legend style={font = \footnotesize,legend cell align=left, align=left, draw=white!15!black},
	}
}
\usepackage[]{algorithm2e}

\usepackage[noadjust]{cite}
\makeatletter
\newcommand{\citecomment}[2][]{\citen{#2}#1\citevar}
\newcommand{\citeone}[1]{\citecomment{#1}}
\newcommand{\citetwo}[2][]{\citecomment[,~#1]{#2}}
\newcommand{\citevar}{\@ifnextchar\bgroup{;~\citeone}{\@ifnextchar[{;~\citetwo}{]}}}
\newcommand{\citefirst}{\@ifnextchar\bgroup{\citeone}{\@ifnextchar[{\citetwo}{]}}}

\makeatother

\pgfplotsset{compat=newest}

\newcommand{\comment}[1]{}

\IEEEoverridecommandlockouts

\begin{document}
		\title{
			Cosine Annealing Optimized Denoising Diffusion Error Correction Codes}
		\author{
		Congyang Ou, Xiaojing Chen, Wan Jiang*\\
			\thanks{
				This research is supported by the National Natural
				Science Foundation of China (No. 62302003.) and the Anhui University Collaborative Innovation Project (No. GTTX-2022-055.) \textit{(Corresponding author: Wan Jiang.)}
			}%
			\thanks{
				Congyang Ou ang Xiaojing Chen are  with the School of Internet, Anhui University, Hefei 230039, China (e-mail: ocy2024@163.com, chenxiaojing0909@ahu.edu.cn).
			}%
			\thanks{
			   Wan Jiang is with the School of Computer Science and Information Engineering, Hefei University of Technology, Hefei 230601, China (e-mail: xjiangw000@gmail.com).
			}
		} 
		
		\maketitle
		\begin{abstract}
	   To address the issue of increased bit error rates during the later stages of linear search in denoising diffusion error correction codes, we propose a novel method that optimizes denoising diffusion error correction codes (ECC) using cosine annealing. In response to the challenge of decoding long codewords, the proposed method employs a variance adjustment strategy during the reverse diffusion process, rather than maintaining a constant variance. By leveraging cosine annealing, this method effectively lowers the bit error rate and enhances decoding efficiency. This letter extensively validates the approach through experiments and demonstrates significant improvements in bit error rate reduction and iteration efficiency compared to existing methods. This advancement offers a promising solution for improving ECC decoding performance, potentially impacting secure digital communication practices.
			
		\end{abstract}
		\section{Introduction}
	     
	    With the rapid development of deep learning, neural networks (NNs) have emerged as a promising solution to complex challenges in digital communication such as channel estimation, equalization, encoding and decoding, leveraging nonlinear processing and efficient hardware implementations \cite{ref1}. However, decoders utilizing generic NN architectures encounter difficulties when decoding long codewords \cite{ref2,ref3,ref4,ref5}. Due to the exponential growth of the codeword space, they often require costly parameter tuning or graph preprocessing to ensure reliability and reduce the risk of overfitting \cite{ref5}.
	     
	     Diffusion Probabilistic Models were first introduced by \cite{ref6}, \cite{ref7} presented DDPM, a generative model for image generation employing a diffusion process. \cite{ref8} utilized a technique based on stochastic differential equations to enhance the quality of sampling. \cite{ref9} proposed DDIMS to accelerate the sampling speed. \cite{ref10} discovered that learning the variance of the inverse diffusion process could reduce the sampling by an order of magnitude in the forward pass, with almost no difference in sample quality. \cite{ref11} introduced the estimation of optimal covariance and its correction for a given imperfect mean through learning conditional expectations, employed in the process of reverse diffusion.
	     
	     Recently, ECCT \cite{ref12} was introduced as a soft decoding approach extending the Transformer architecture to linear codes of arbitrary block lengths. The interaction between algebraic code and bits is incorporated into the model through an adapted masked self-attention module. Subsequently, DDECC \cite{ref13} was proposed as an extension of ECCT, introducing a diffusion process suitable for decoding settings. This neural decoder optimizes based on the quantity of parity errors and introduces a linear search method based on syndromes. Furthermore, DQEC \cite{ref14} has effectively reduced the error rate in quantum computing by developing a novel end-to-end deep quantum error decoder, demonstrating the powerful potential of utilizing deep learning to optimize quantum error correction codes. Its performance surpasses that of existing neural and classical decoders.
         
         In the process of replicating the current state-of-the-art framework \cite{ref13}, we observed that its sampling speed was excessively fast, which could paradoxically  increase the final bit error rate. Therefore, in this letter, we design the following methods based on cosine annealing: during the sampling process, we employ the designed cosine variance instead of fixing it at 0.01; we attempt to use cosine variance in the loss function during the training process; and we integrate the use of cosine variance with linear search during the sampling process.

         Through experimental documentation and comparison with the methods in \cite{ref13}, we find that using cosine variance during the sampling process can effectively reduce the bit error rate. Moreover, integrating the use of cosine variance with linear search during the sampling process can further improve sampling efficiency based on linear search, and reduce the number of syndrome iterations to zero.
          \section{Cosine Annealing}
          In diffusion models, the variance within the reverse diffusion process has always been a thought-provoking issue. As defined in the article \cite{ref13} DDECC, the reverse diffusion process is described as follows:
           \begin{equation}
          	x_{t-1}=x_{t} -\lambda \frac{\sqrt{\bar{\beta _{t} } }\beta _{t}  }{\bar{\beta _{t} }\beta _{t} } \hat{\varepsilon}  .
          \label{linesample}
          \end{equation}
          
          In the above equation, $\hat{\varepsilon} $ represents additive noise, and $\beta _{t}$ represents the noise variance. In the approach  DDECC,  $\beta _{t}$ is to fixed at 0.01.  $\lambda$ is a parameter for linear search; if linear search is not employed,  $\lambda$  is set to 1; otherwise, $\lambda$  is determined as the parameter that minimizes the bit error rate.

          Employing linear search significantly enhances the efficiency of sampling. However, experiments have revealed that as the number of iterations increases, it is often observed that the bit error rate of methods employing linear search can be inferior to those not utilizing linear search, as illustrated in the following figure:
\begin{figure}[h]
	\centering
		\includegraphics[width=\columnwidth]{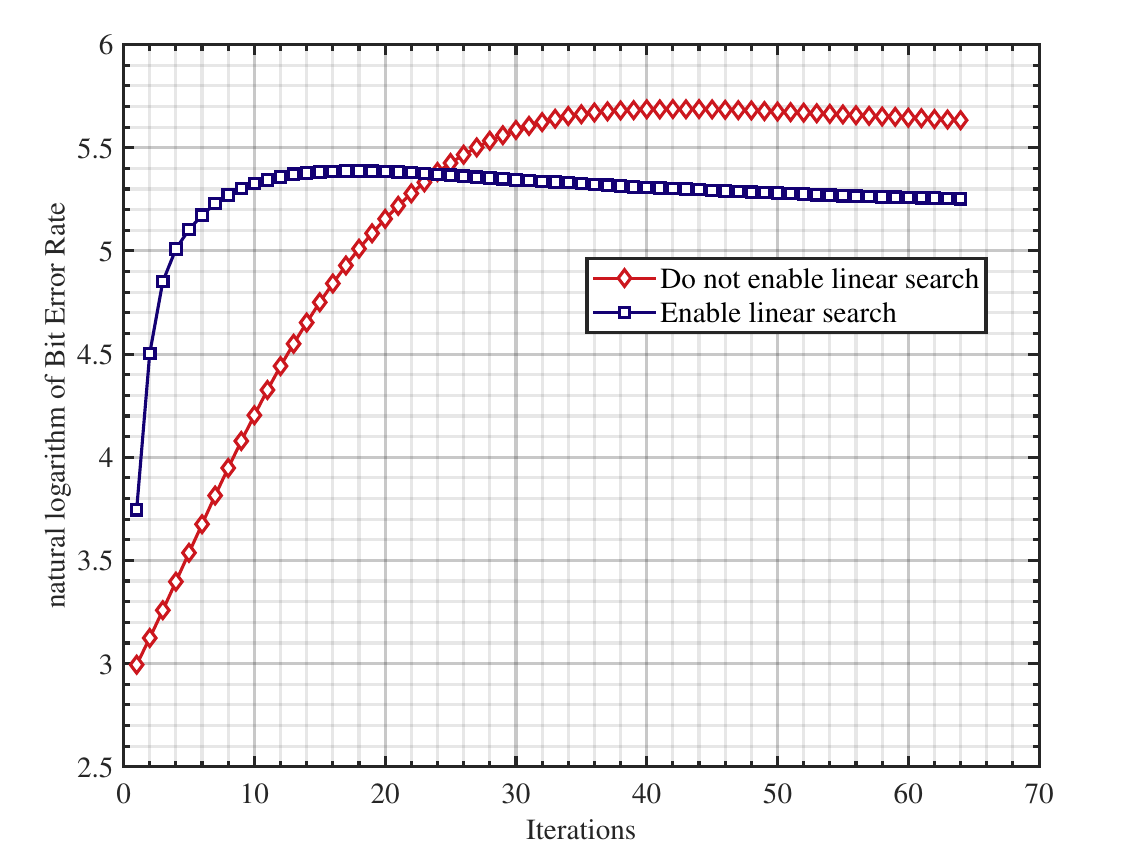}
	
	\caption{
		For Polar (128, 64) with 	N=2 and D=32, the comparison of the negative logarithm of the bit error rate with linear search enabled versus disabled is presented (the higher this value, the better the performance).}
	\label{fig1}
\end{figure}
       
        The graph clearly demonstrates that initially, methods with linear search activated possess a significant advantage. However, as the number of iterations increases, the negative logarithm of the bit error rate may even decrease, and performance becomes inferior to that of methods without linear search. Through our experiments, this has been observed not as an isolated case but rather a common phenomenon across the majority of codes. Consequently, this motivates us to design a method whose rate of decline is between that of linear search and no linear search. This approach not only avoids the slow decline associated with disabling linear search but also moderates the decrease observed with linear search. Thereby it effectively reduces the bit error rate.

         According to \cite{ref10} IDDPM, linear scheduling results in a quicker reduction to zero and more significant information degradation. Therefore, a cosine noise schedule was designed. We have decided to use such a cosine noise variance to serve a moderating function in the decline. The specific noise variance is defined as follows:
         \begin{equation}
         \beta_i = \min\left(1 - \frac{\cos^2\left(\frac{\frac{i + 1}{N} + 0.008}{1.008} \cdot \frac{\pi}{2}\right)}{\cos^2\left(\frac{\frac{i}{N} + 0.008}{1.008} \cdot \frac{\pi}{2}\right)}, \beta_{\text{max}}\right).
         \label{beta_cos}
         \end{equation}    
          Here, $i$ represents the current time step $i=0,1, ...,N-1$,  $ N $ is the total number of time diffusion steps, and $\beta_{max}$  is the maximum allowable value for $\beta_{t}$, set to 0.999 to prevent singularities near the end of the diffusion process when $t=T$. Additionally, an offset of 0.008 is introduced to prevent $\beta_{t}$ from being too small near  $t=0$. The set of noise variances constructed above is henceforth uniformly represented by $\beta_{cos}$ in the subsequent text.
         
          Therefore, we use the formula shown below to sample the reverse diffusion process.
           \begin{equation}
          	x_{t-1}=x_{t} - \frac{\sqrt{{\bar{\beta}}_{cos} }\beta _{cos}  }{{\bar{\beta}}_{cos}\beta _{cos} } \hat{\varepsilon}  .
          \label{cossample}
          \end{equation}
          It is important to note that the sampling process here does not initiate linear search; it merely utilizes cosine variance.
          
          The negative logarithm of the bit error rate brought about by linear search and our cosine variance is presented in Table \ref{tab1} as shown.
          As can be seen from Table \ref{tab1}, the use of cosine annealing results in a reduction in the negative logarithm of the bit error rate for the vast majority of codewords.
          
          \begin{table*}[htbp]
          	\centering
          	\caption{
          		The comparison of the maximum negative logarithm of the Bit Error Rate during the iterative processes of linear search and cosine annealing is displayed in the table, where, from top to bottom $d=32.64.128$. $4,5,6$ represent  EbN0 (non-normalized SNR).}
          	\begin{tabular}{rrrrrrrr||rrrrrrrr}
          		    \toprule
          		\multicolumn{1}{l}{Method} & \multicolumn{3}{c}{Line N=2} &       & \multicolumn{3}{c}{Line N=6} &       & \multicolumn{3}{c}{Cosine N=2} &       & \multicolumn{3}{c}{Cosine N=6} \\
          		\cmidrule{2-4}\cmidrule{6-8}\cmidrule{10-12}\cmidrule{14-16}          & \multicolumn{1}{c}{4} & \multicolumn{1}{c}{5} & \multicolumn{1}{c}{6} &       & \multicolumn{1}{c}{4} & \multicolumn{1}{c}{5} & \multicolumn{1}{c||}{6} &       & \multicolumn{1}{c}{4} & \multicolumn{1}{c}{5} & \multicolumn{1}{c}{6} &       & \multicolumn{1}{c}{4} & \multicolumn{1}{c}{5} & \multicolumn{1}{c}{6} \\
          		\midrule
          		\multicolumn{1}{r}{\multirow{3}[2]{*}{Polar(64,32)}} & 6.086 & 8.258 & 11.052 &       & 6.556 & 8.746 & 11.821 &       & 6.294 & 8.549 & 11.422 &       & 6.652 & 9.090 & 12.242 \\
          		& 6.353 & 8.667 & 11.483 &       & 6.892 & 9.577 & 12.826 &       & 6.576 & 8.825 & 11.634 &       & 7.131 & 9.601 & 13.076 \\
          		& 6.550 & 8.892 & 12.065 &       & 7.124 & 9.611 & 13.131 &       & 6.687 & 9.026 & 12.162 &       & 7.238 & 9.770 & 13.167 \\
          		\midrule
          		\multicolumn{1}{r}{\multirow{3}[2]{*}{Polar(64,48)}} & 5.633 & 7.640 & 10.144 &       & 5.969 & 8.142 & 10.997 &       & 5.684 & 7.806 & 10.299 &       & 6.078 & 8.175 & 10.911 \\
          		& 5.737 & 7.828 & 10.254 &       & 6.079 & 8.266 & 11.076 &       & 5.829 & 7.866 & 10.693 &       & 6.125 & 8.35  & 11.189 \\
          		& 5.809 & 7.966 & 10.610 &       & 6.020 & 8.206 & 10.893 &       & 5.934 & 8.084 & 10.700 &       & 6.115 & 8.33  & 10.909 \\
          		\midrule
          		\multicolumn{1}{r}{\multirow{3}[2]{*}{Polar(128,64)}} & 5.387 & 7.803 & 10.718 &       & 6.441 & 9.358 & 12.746 &       & 5.688 & 8.087 & 11.141 &       & 6.678 & 9.598 & 13.268 \\
          		& 5.983 & 8.740 & 11.719 &       & 7.474 & 10.818 & 14.611 &       & 6.273 & 9.084 & 12.278 &       & 7.654 & 11.022 & 14.755 \\
          		& 6.716 & 9.504 & 12.680 &       & 8.246 & 12.058 & 15.946 &       & 6.936 & 10.072 & 12.232 &       & 8.225 & 11.845 & 14.512 \\
          		\midrule
          		\multicolumn{1}{r}{\multirow{3}[2]{*}{Polar(128,86)}} & 5.581 & 7.682 & 10.410 &       & 6.479 & 9.151 & 12.688 &       & 5.768 & 7.937 & 10.648 &       & 6.804 & 9.565 & 12.935 \\
          		& 6.036 & 8.282 & 11.111 &       & 7.101 & 9.990 & 13.780 &       & 6.146 & 8.388 & 11.164 &       & 7.367 & 10.434 & 14.305 \\
          		& 6.309 & 8.721 & 11.776 &       & 7.692 & 11.280 & 16.472 &       & 6.381 & 8.733 & 11.892 &       & 7.874 & 11.381 & 15.461 \\
          		\midrule
          		\multicolumn{1}{r}{\multirow{3}[2]{*}{Polar(128,96)}} & 5.642 & 7.906 & 10.626 &       & 6.446 & 9.327 & 12.698 &       & 5.712 & 7.982 & 10.867 &       & 6.636 & 9.425 & 12.688 \\
          		& 5.955 & 8.442 & 11.423 &       & 6.782 & 9.582 & 12.887 &       & 6.038 & 8.464 & 11.496 &       & 6.906 & 9.627 & 13.102 \\
          		& 6.297 & 8.910 & 11.951 &       & 7.187 & 10.182 & 13.214 &       & 6.434 & 9.027 & 12.187 &       & 7.331 & 10.434 & 13.263 \\
          		\midrule
          		\multicolumn{1}{r}{\multirow{3}[2]{*}{LDPC(49,24)}} & 5.227 & 7.381 & 10.051 &       & 5.820 & 8.269 & 11.446 &       & 5.296 & 7.348 & 10.113 &       & 5.935 & 8.306 & 11.632 \\
          		& 5.295 & 7.441 & 10.317 &       & 5.887 & 8.360 & 11.547 &       & 5.312 & 7.428 & 10.333 &       & 5.971 & 8.514 & 11.655 \\
          		& 5.352 & 7.367 & 10.136 &       & 5.846 & 8.331 & 11.726 &       & 5.358 & 7.525 & 10.322 &       & 5.933 & 9.406 & 11.948 \\
          		\midrule
          		\multicolumn{1}{r}{\multirow{3}[2]{*}{LDPC(121,60)}} & 4.553 & 7.019 & 10.519 &       & 5.325 & 8.476 & 13.529 &       & 4.593 & 7.055 & 10.804 &       & 5.356 & 8.731 & 13.769 \\
          		& 4.567 & 6.982 & 10.899 &       & 5.380 & 8.772 & 13.850 &       & 4.624 & 7.168 & 10.968 &       & 5.478 & 8.968 & 14.507 \\
          		& 4.582 & 7.077 & 10.869 &       & 5.489 & 8.984 & 14.302 &       & 4.624 & 7.148 & 10.986 &       & 5.583 & 9.008 & 14.647 \\
          		\midrule
          		\multicolumn{1}{r}{\multirow{3}[2]{*}{LDPC(121,70)}} & 5.505 & 8.383 & 12.440 &       & 6.631 & 10.444 & 16.009 &       & 5.635 & 8.604 & 12.472 &       & 6.661 & 10.404 & 16.557 \\
          		& 5.512 & 8.348 & 12.412 &       & 6.662 & 10.825 & 16.339 &       & 5.662 & 8.613 & 12.838 &       & 6.807 & 11.045 & 16.944 \\
          		& 5.550 & 8.504 & 12.679 &       & 6.823 & 10.841 & 16.381 &       & 5.682 & 8.665 & 12.931 &       & 6.988 & 11.166 & 17.225 \\
          		\midrule
          		\multicolumn{1}{r}{\multirow{3}[2]{*}{LDPC(121,80)}} & 6.206 & 9.241 & 13.149 &       & 7.473 & 11.727 & 17.130 &       & 6.358 & 9.511 & 13.693 &       & 7.612 & 11.829 & 16.871 \\
          		& 6.286 & 9.362 & 13.594 &       & 7.619 & 11.963 & 17.330 &       & 6.404 & 9.667 & 13.805 &       & 7.840 & 12.270 & 17.918 \\
          		& 6.326 & 9.675 & 13.727 &       & 7.705 & 12.305 & 18.613 &       & 6.407 & 9.753 & 13.983 &       & 7.958 & 12.419 & 18.349 \\
          		\midrule
          		\multicolumn{1}{r}{\multirow{3}[2]{*}{MacKay(96,48)}} & 6.332 & 8.686 & 11.787 &       & 7.813 & 11.251 & 15.091 &       & 6.352 & 8.872 & 11.792 &       & 7.996 & 11.309 & 15.535 \\
          		& 6.294 & 8.807 & 12.026 &       & 8.010 & 11.582 & 15.49 &       & 6.434 & 9.094 & 12.020 &       & 8.130 & 11.873 & 15.665 \\
          		& 6.328 & 9.016 & 12.034 &       & 8.107 & 11.926 & 15.807 &       & 6.485 & 9.041 & 12.243 &       & 8.350 & 12.183 & 15.830 \\
          		\midrule
          		\multicolumn{1}{r}{\multirow{3}[2]{*}{CCSDS(128,64)}} & 5.755 & 8.534 & 12.074 &       & 7.485 & 11.751 & 16.528 &       & 5.855 & 8.797 & 12.218 &       & 7.624 & 11.679 & 17.879 \\
          		& 5.771 & 8.466 & 12.018 &       & 7.777 & 11.984 & 17.603 &       & 6.055 & 8.897 & 12.389 &       & 7.804 & 12.409 & 17.415 \\
          		& 5.810 & 8.836 & 12.386 &       & 7.943 & 12.263 & 16.796 &       & 6.056 & 9.067 & 13.099 &       & 8.107 & 12.537 & 17.307 \\
          		\midrule
          		\multicolumn{1}{r}{\multirow{3}[2]{*}{BCH(31,16)}} & 4.487 & 5.839 & 7.714 &       & 4.495 & 5.929 & 7.875 &       & 4.595 & 6.034 & 7.984 &       & 4.769 & 6.232 & 8.252 \\
          		& 4.674 & 6.182 & 8.301 &       & 4.757 & 6.319 & 8.258 &       & 4.827 & 6.376 & 8.385 &       & 4.863 & 6.383 & 8.437 \\
          		& 4.966 & 6.596 & 8.821 &       & 5.112 & 6.735 & 8.942 &       & 5.195 & 6.858 & 9.094 &       & 5.262 & 6.958 & 9.019 \\
          		\midrule
          		\multicolumn{1}{r}{\multirow{3}[2]{*}{BCH(63,36)}} & 4.686 & 6.383 & 8.769 &       & 4.953 & 6.815 & 9.402 &       & 4.767 & 6.474 & 8.777 &       & 5.048 & 6.881 & 9.513 \\
          		& 4.812 & 6.594 & 8.923 &       & 5.142 & 7.114 & 9.860 &       & 4.900 & 6.726 & 9.321 &       & 5.244 & 7.224 & 9.768 \\
          		& 5.190 & 7.183 & 10.128 &       & 5.204 & 7.216 & 9.840 &       & 5.313 & 7.309 & 10.180 &       & 5.320 & 7.362 & 9.900 \\
          		\midrule
          		\multicolumn{1}{r}{\multirow{3}[2]{*}{BCH(63,45)}} & 5.208 & 7.246 & 10.13 &       & 5.582 & 7.858 & 10.883 &       & 5.260 & 7.333 & 10.294 &       & 5.684 & 7.903 & 10.982 \\
          		& 5.311 & 7.451 & 10.313 &       & 5.683 & 7.982 & 11.144 &       & 5.363 & 7.528 & 10.317 &       & 5.777 & 8.238 & 11.324 \\
          		& 5.449 & 7.583 & 10.497 &       & 5.767 & 8.169 & 11.366 &       & 5.525 & 7.775 & 10.667 &       & 5.826 & 8.481 & 11.762 \\
          		\bottomrule
          	\end{tabular}%
          	\label{tab1}%
          \end{table*}%
         \section{Cosine Iterative Optimization}
         As mentioned in the previous section, employing the cosine annealing method can effectively reduce the bit error rate. However, because it moderates the sampling intensity of linear codes, it leads to a greater number of syndrome reductions to zero compared to the linear search method, positioning it between the scenarios of linear search being enabled and disabled. Consequently, there is a need to design a method that can reduce the number of syndrome reductions to zero even more than the linear search method, while also hoping it can lower the bit error rate.

         In \cite{ref13}, the loss function during the training process is as follows: 
           \begin{equation}\label{linetrain}
         	\mathcal{L}(\theta) = -\mathbb{E}_{t,x_0, \varepsilon}\log\Big(\epsilon_{\theta}(x_{0}+\sqrt{{\bar{\beta}}_{t}}\varepsilon, {e_{t}}), \tilde{\varepsilon}_{b}\Big)\,.
         \end{equation}
        Here, ${e_{t}}$ represents parity check errors, $\epsilon_{\theta}$ refers to the parameters of the decoder obtained through training, and $\tilde{\varepsilon}_{b}$ denotes the predicted multiplicative noise.
        
        As previously mentioned, cosine annealing can play a moderating role. Therefore, this letter decides to first apply the cosine noise variance to the loss function of the training process, as shown in the following formula:
           \begin{equation}\label{costrain}
        	\mathcal{L}(\theta) = -\mathbb{E}_{t,x_0, \varepsilon}\log\Big(\epsilon_{\theta}(x_{0}+\sqrt{\pmb{{\bar{\beta}}_{cos}}}\varepsilon, {e_{t}}), \tilde{\varepsilon}_{b}\Big)\,.
        \end{equation}
        In subsequent sections, the loss function as described in Eq. \ref{costrain} is referred to as ``cosine training," while the approach described in Eq. \ref{linetrain} is referred to as ``linear training."

        Besides, due to the insufficient sampling efficiency of cosine annealing and the overly high sampling efficiency of linear search, a consideration is made to blend the two to enhance sampling efficiency, as shown in the following formula:
        \begin{equation}
        	x_{t-1}=x_{t} - \lambda\frac{\sqrt{{\bar{\beta}}_{cos} }\beta _{cos}  }{{\bar{\beta}}_{cos}\beta _{cos} } \hat{\varepsilon}  .
        \label{intergrated sample}
        \end{equation}
        Differing from \cite{ref13}, here $\lambda$ is obtained by solving the problem described below:
        \begin{equation}
        	\begin{aligned}
        		\lambda^{*} = argmin_{\lambda \in \mathbb{R}^{+}} \|s\Big(x_{t}-\lambda \frac{\sqrt{\bar{\beta}_{cos}}\beta_{cos}}{\bar{\beta}_{cos}+\beta_{cos}}\hat{\varepsilon}\Big)\|_{1}.
        	\end{aligned}
        	\label{eq:ddecc_LS}
        \end{equation}
        Here, $s(y)$ represents the binary code syndrome, that is, the number of bit errors.
        In the subsequent text, the method employing Eq. \ref{linesample} is referred to as linear sampling, the method employing Eq. \ref{cossample} is referred to as cosine sampling, and the method employing Eq. \ref{intergrated sample} is referred to as integrated sampling.
       \section{experiments} 
       We employ the state-of-the-art DDECC architecture as presented in \cite{ref13}. In this architecture, the model's capability is defined by the selected embedding dimension 
       $d$ and the number of self-attention layers $N$. A discrete grid search for the linear search parameter $\lambda$ samples 20 points uniformly across  $I=[1,20]$ to find the optimal step size.

       Optimization is performed using the Adam \cite{ref15} optimizer, with each mini-batch containing 128 samples, for a total of 2000 epochs, each consisting of 1000 mini-batches. Regarding noise variance, for linear sampling, the variance is fixed at 0.01. For cosine sampling, the noise variance table is calculated using Eq. \ref{beta_cos}. For integrated sampling, the cosine variance is still used.
       We initialize the learning rate at $10^{-4} $ and employ a cosine decay scheduler to reduce it to $5.10^{-6} $ by the end of training. No warm-up phase is utilized \cite{ref16}.
       
       To evaluate our method, we trained using four architectures as shown in Table \ref{tab2}, employing codewords consistent with those used in \cite{ref13} to assess the efficacy of our approach.
       It is important to note that Method 1 in Table \ref{tab2} is the approach proposed in \cite{ref13}, while the remaining three methods are entirely new approaches that we have introduced.
       
       The proposed architecture is defined solely by the number of encoder layers 
       $N$ and the embedding dimension 
       $d$. During the training process, we evaluate six different architectures with $N={2,6}$ and  $d={32,64,128}$, and train them using the four architectures as shown in Table \ref{tab2}.
\begin{table}[htbp]
	\centering
	\caption{methods}
	\begin{tabular}{ccccc}
		\toprule
		\multicolumn{5}{c}{\textcolor{black}{Training and Sampling Methods}} \\
		\midrule
		1     & \multicolumn{2}{c}{linear training} & \multicolumn{2}{c}{linear sampling} \\
		2     & \multicolumn{2}{c}{linear training} & \multicolumn{2}{c}{cosine sampling} \\
		3     & \multicolumn{2}{c}{linear training} & \multicolumn{2}{c}{integrated sampling} \\
		4     & \multicolumn{2}{c}{cosine training} & \multicolumn{2}{c}{linear sampling} \\
		\bottomrule
	\end{tabular}
	\label{tab2}
\end{table}
      
      We report the Bit Error Rate (BER) for different values of the normalized signal-to-noise ratio (Eb/N0). We follow the testing benchmarks outlined in \cite{ref13}. During the testing period, our decoder decoded at least $10^{5} $  random codewords to achieve a minimum of 500 frames with errors at each SNR value. All results were obtained through experimental procedures.
      
     The negative logarithm of the Bit Error Rate is recorded in Table \ref{tab1}. It shows that cosine annealing (Method 2) reduces the Bit Error Rate more effectively than linear search (Method 1), though its iteration efficiency is lower. Thus, other training methods were employed, and Table \ref{tab3} lists the iterations needed for each codeword's syndrome to converge to zero, with the best method indicated by a subscript. Table \ref{tab3} clearly demonstrates that our method, especially Method 3, achieves better iterative performance on most codewords. The specific iterative trends are depicted in the figure below:
    \begin{figure}[h]
    	\centering
    	\includegraphics[width=\columnwidth]{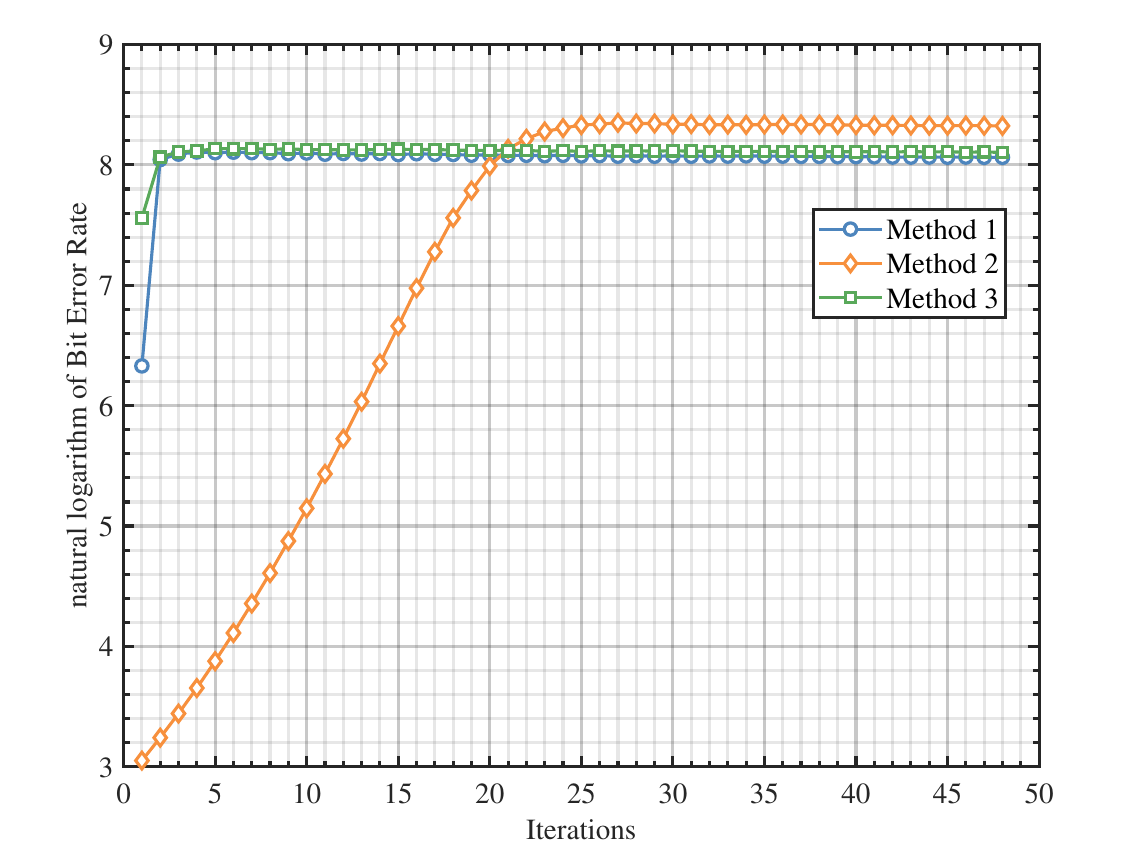}
    	
    	\caption{
    		For MACKAY(96,48) with 	N=6 and D=128, the comparison of the negative logarithm of the bit error rate with different methods(the higher this value, the better the performance).}
         \label{fig2}
    \end{figure}
  
  As depicted in Fig. \ref{fig2}, Method 2, which utilizes cosine annealing, shows the best performance in reducing the bit error rate. Although Methods 3 and 1 achieve similar final bit error rates, they are more effective in the initial stages, enabling more efficient sampling and reducing the number of times the syndrome iterates to zero. It is important to emphasize that this is not an isolated case; most codewords conform to this pattern.
  
     \begin{table}[htbp]
     	\centering
     	\caption{
     		Comparison of the average number of iterations required for the syndrome of different codewords to converge to zero. The subscript in the lower right corner indicates which method yields the best iterative performance. From top to bottom, d successively represents 32, 64, 128.}
     	\footnotesize 
     	\begin{tabular}{@{}lcccccc@{}}
     		\toprule
     	Method & \multicolumn{3}{c}{N=2} & \multicolumn{3}{c}{N=6} \\
     	\cmidrule(lr){2-4} \cmidrule(lr){5-7}
     	& 4 & 5 & 6 & 4 & 5 & 6 \\
     	\midrule
     	\multirow{3}{*}{Polar(64,32)} & $1.535_{3}$ & $1.019_{3}$ & $0.785_{3}$ & $1.082_{3}$ & $0.927_{3}$ & $0.775_{3}$ \\
     	& $1.375_{3}$ & $0.980_{3}$ & $0.781_{3}$ & $1.025_{3}$ & $0.918_{3}$ & $0.774_{3}$ \\
     	& $1.296_{3}$ & $0.960_{3}$ & $0.777_{3}$ & $1.012_{3}$ & $0.918_{3}$ & $0.744_{3}$ \\
     	\midrule
     	\multirow{3}{*}{Polar(64,48)} & $1.397_{1}$ & $0.703_{1}$ & $0.380_{1}$ & $0.953_{3}$ & $0.621_{3}$ & $0.373_{3}$ \\
     	& $1.265_{3}$ & $0.680_{3}$ & $0.376_{3}$ & $0.940_{3}$ & $0.622_{3}$ & $0.372_{3}$ \\
     	& $1.197_{1}$ & $0.659_{3}$ & $0.376_{3}$ & $0.953_{3}$ & $0.628_{3}$ & $0.372_{3}$ \\
     	\midrule
     	\multirow{3}{*}{Polar(128,64)} & $4.602_{3}$ & $1.601_{3}$ & $1.019_{3}$ & $1.377_{3}$ & $1.023_{3}$ & $0.952_{3}$ \\
     	& $3.240_{3}$ &  $1.323_{3}$ & $0.984_{3}$ & $1.458_{3}$ & $1.040_{3}$ & $0.953_{3}$  \\
     	& $2.432_{3}$ & $1.180_{3}$ & $0.967_{3}$ & $1.108_{3}$ & $0.999_{3}$ & $0.951_{3}$ \\
     	\midrule
     	\multirow{3}{*}{Polar(128,86)} & $3.539_{3}$ & $1.360_{3}$ & $0.788_{3}$ & $1.421_{3}$ & $0.960_{3}$ & $0.737_{3}$ \\
     	& $3.024_{3}$ & $1.246_{3}$ & $0.766_{3}$ & $1.215_{3}$ & $0.935_{3}$ & $0.736_{3}$ \\
     	& $2.453_{3}$ & $1.118_{3}$ & $0.751_{3}$ & $1.114_{3}$ & $0.926_{3}$ & $0.736_{3}$ \\
     	\midrule
     	\multirow{3}{*}{Polar(128,96)} & $3.639_{3}$ & $1.219_{3}$ & $0.629_{3}$ & $1.341_{1}$ & $0.871_{3}$ & $0.608_{3}$ \\
     	& $2.749_{3}$ & $1.035_{3}$ & $0.618_{3}$ & $1.205_{3}$ & $0.863_{3}$ & $0.607_{3}$ \\
     	& $1.972_{3}$ & $0.945_{3}$ & $0.612_{3}$ & $1.105_{3}$ & $0.856_{3}$ & $0.607_{3}$ \\
     	\midrule
     	\multirow{3}{*}{LDPC(49,24)} & $2.213_{3}$ & $1.021_{3}$ & $0.712_{3}$ & $1.436_{3}$ & $0.910_{3}$ & $0.700_{3}$ \\
     	& $2.100_{3}$ & $1.020_{3}$ & $0.711_{3}$ & $1.342_{4}$ & $0.897_{4}$ & $0.698_{4}$ \\
     	& $2.034_{3}$ & $1.027_{3}$ & $0.710_{3}$ & $1.455_{3}$ & $0.901_{3}$ & $0.699_{3}$ \\
     	\midrule
     	\multirow{3}{*}{LDPC(121,60)} & $9.981_{3}$ & $2.015_{3}$ & $0.985_{3}$ & $4.468_{3}$ & $1.138_{3}$ & $0.944_{3}$ \\
     	& $9.844_{3}$ & $1.990_{3}$ & $0.979_{3}$ & $4.329_{3}$ & $1.121_{3}$ & $0.944_{3}$ \\
     	& $9.612_{3}$ & $1.949_{3}$ & $0.977_{3}$ & $3.888_{3}$ & $1.092_{3}$ & $0.944_{3}$ \\
     	\midrule
     	\multirow{3}{*}{LDPC(121,70)} & $4.484_{3}$ & $1.259_{3}$ & $0.865_{3}$ & $1.881_{3}$ & $0.991_{3}$ & $0.856_{3}$ \\
     	& $4.558_{3}$ & $1.237_{3}$ & $0.864_{3}$ & $1.820_{3}$ & $0.982_{3}$ & $0.856_{3}$ \\
     	& $4.149_{3}$ & $1.203_{3}$ & $0.863_{3}$ & $1.683_{3}$ & $0.978_{3}$ & $0.856_{3}$ \\
     	\midrule
     	\multirow{3}{*}{LDPC(121,80)} & $2.755_{3}$ & $1.032_{3}$ & $0.737_{3}$ & $1.298_{3}$ & $0.923_{3}$ & $0.733_{3}$ \\
     	& $2.525_{3}$ & $1.019_{3}$ & $0.736_{3}$ & $1.273_{3}$ & $0.922_{3}$ & $0.733_{3}$ \\
     	& $2.462_{3}$ & $0.999_{3}$ & $0.736_{3}$ & $1.215_{3}$ & $0.920_{3}$ & $0.733_{3}$ \\
     	\midrule
     	\multirow{3}{*}{MacKay(96,48)} & $2.657_{1}$ & $1.167_{1}$ & $0.904_{3}$ & $1.209_{3}$ & $0.982_{3}$ & $0.892_{3}$ \\
     	& $2.581_{1}$ & $1.156_{1}$ & $0.736_{3}$ & $1.149_{3}$ & $0.980_{3}$ & $0.892_{3}$ \\
     	& $2.450_{3}$ & $1.163_{3}$ & $0.901_{3}$ & $1.145_{3}$ & $0.978_{3}$ & $0.892_{3}$ \\
     	\midrule 
     	\multirow{3}{*}{CCSDS(128,64)} & $4.803_{3}$ & $1.423_{3}$ & $0.970_{3}$ & $1.421_{3}$ & $1.011_{3}$ & $0.949_{3}$ \\
     	& $4.292_{3}$ & $1.310_{3}$ & $0.966_{3}$ & $1.331_{3}$ & $0.998_{3}$ & $0.949_{3}$ \\
     	& $4.124_{3}$ & $1.281_{3}$ & $0.964_{3}$ & $1.261_{3}$ & $0.996_{3}$ & $0.949_{3}$ \\
     	\midrule
     	\multirow{3}{*}{BCH(31,16)} & $1.111_{4}$ & $0.741_{4}$ & $0.496_{4}$ & $0.854_{4}$ & $0.677_{4}$ & $0.487_{4}$ \\
     	& $1.011_{4}$ & $0.706_{4}$ & $0.493_{4}$ & $0.841_{4}$ & $0.677_{4}$ & $0.487_{4}$ \\
     	& $0.878_{3}$ & $0.680_{3}$ & $0.488_{3}$ & $0.831_{4}$ & $0.673_{4}$ & $0.487_{4}$ \\
     	\midrule
     	\multirow{3}{*}{BCH(63,36)} & $3.736_{1}$ & $1.467_{1}$ & $0.733_{1}$ & $2.816_{1}$ & $1.182_{3}$ & $0.679_{3}$ \\
     	& $3.358_{1}$ & $1.363_{1}$ & $0.712_{3}$ & $2.422_{1}$ & $1.088_{1}$ & $0.669_{1}$ \\
     	& $2.601_{1}$ & $1.150_{1}$ & $0.677_{1}$ & $2.319_{1}$ & $1.055_{1}$ & $0.665_{3}$ \\
     	\midrule
     	\multirow{3}{*}{BCH(63,45)} & $2.065_{1}$ & $0.856_{1}$ & $0.435_{1}$ & $1.480_{1}$ & $0.732_{1}$ & $0.422_{3}$ \\
     	& $1.924_{1}$ & $0.822_{1}$ & $0.429_{1}$ & $1.387_{1}$ & $0.712_{1}$ & $0.420_{3}$ \\
     	& $1.750_{1}$ & $0.784_{1}$ & $0.428_{1}$ & $1.349_{1}$ & $0.709_{1}$ & $0.419_{3}$ \\

     	\bottomrule
     	\end{tabular}%
     	\label{tab3}
     \end{table}
     \section{conclusion}
    Our designed methods all achieve better results on the existing basis. Among them, Method 2 can minimize the bit error rate to the greatest extent, but its iteration efficiency is not as good as the previously best method. On the other hand, Method 3 has better iteration efficiency than the previously best method, but it does not have a significant advantage in terms of bit error rate. Instead, it effectively reduces the number of syndrome iterations to zero.
       \nocite{*}
       \bibliographystyle{IEEEtran}
       \bibliography{reference.bib}

\end{document}